\documentclass[aps,a4paper,twocolumn,nofootinbib,superscriptaddress]{revtex4}
\usepackage{amsmath,amssymb,color,dsfont}

\usepackage{pdfpages}

\newcommand{\trace}[1]{\textrm{Tr}\left(#1\right)}

\newcommand{\abs}[1]{\left|#1\right|}

\newcommand{\integ}[1]{\ensuremath{\int \!\! \mathrm{d}#1 \,}}
\newcommand{\integlim}[3]{\ensuremath{\int_{#1}^{#2} \!\!\! \mathrm{d}#3 \,}}

\newcommand{\mean}[1]{\ensuremath{\left\langle #1 \right\rangle}}

\newcommand{\prob}[1]{\textrm{Pr} \! \left(#1\right)}

\newcommand{\nbar}{\ensuremath{\bar{n}}}

\newcommand{\xzp}{\ensuremath{x_0}}

\newcommand{\omegam}{\ensuremath{\omega_\textrm{M}}}

\newcommand{\bra}[1]{\ensuremath{\left\langle #1 \right|}}
\newcommand{\ket}[1]{\ensuremath{\left| #1 \right\rangle}}

\newcommand{\braket}[2]{\ensuremath{\left\langle #1 | #2 \right\rangle}}
\newcommand{\ketbra}[2]{\ensuremath{\left|#1\right\rangle \! \left\langle#2\right|}}

\newcommand{\dis}{\ensuremath{\hat{D}}}
\newcommand{\disd}{\ensuremath{\hat{D}^\dagger}}

\newcommand{\ad}{\ensuremath{\hat{a}^\dagger}}
\newcommand{\bd}{\ensuremath{\hat{b}^\dagger}}

\newcommand{\XM}{\ensuremath{X_\textrm{M}}}

\newcommand{\PL}{\ensuremath{P_\textrm{L}}}

\newcommand{\etal}{\emph{et al}.}
\newcommand{\PRL}[3]{Phys. Rev. Lett.~\textbf{#1}, #2 (#3)}
\newcommand{\PRA}[3]{Phys. Rev. A~\textbf{#1}, #2 (#3)}
\newcommand{\PRAR}[3]{Phys. Rev. A~\textbf{#1}, #2(R) (#3)} 
\newcommand{\PRD}[3]{Phys. Rev. D~\textbf{#1}, #2 (#3)}
\newcommand{\PRX}[3]{Phys. Rev. X~\textbf{#1}, #2 (#3)}
\newcommand{\RMP}[3]{Rev. Mod. Phys.~\textbf{#1}, #2 (#3)}
\newcommand{\Nature}[3]{Nature~\textbf{#1}, #2 (#3)}
\newcommand{\NatComm}[3]{Nat. Commun.~\textbf{#1}, #2 (#3)}
\newcommand{\NatNano}[3]{Nat. Nanotechnol.~\textbf{#1}, #2 (#3)}
\newcommand{\NatPhot}[3]{Nature Photon.~\textbf{#1}, #2 (#3)}
\newcommand{\NatPhys}[3]{Nature Phys.~\textbf{#1}, #2 (#3)}
\newcommand{\ActPSlov}[3]{Acta Physica Slovaca~\textbf{#1}, #2 (#3)}
\newcommand{\Annalen}[3]{Ann. Phys. (Berlin)~\textbf{#1}, #2 (#3)}

\newcommand{\EPJD}[3]{Eur. Phys. J. D~\textbf{#1}, #2 (#3)}
\newcommand{\JETP}[3]{JETP Lett.~\textbf{#1}, #2 (#3)}

\newcommand{\JOptB}[3]{J. Opt. B~\textbf{#1}, #2 (#3)}
\newcommand{\NJP}[3]{New J. Phys.~\textbf{#1}, #2 (#3)}
\newcommand{\OX}[3]{Optics Express~\textbf{#1}, #2 (#3)}
\newcommand{\OptComm}[3]{Opt. Commun.~\textbf{#1}, #2 (#3)}

\newcommand{\PLettA}[3]{Physics Letters A~\textbf{#1}, #2 (#3)}
\newcommand{\PNAS}[3]{Proc. Nat. Acad. Sci. USA~\textbf{#1}, #2 (#3)}
\newcommand{\RPG}[3]{Rep. Prog. Phys.~\textbf{#1}, #2 (#3)}
\newcommand{\Science}[3]{Science~\textbf{#1}, #2 (#3)}

\begin{document}

\title{Towards Optomechanical Quantum State Reconstruction of Mechanical Motion}

\author{M. R. Vanner\footnote{Email correspondence: m.vanner@uq.edu.au}}
\affiliation{School of Mathematics and Physics, The University of Queensland, Brisbane, Queensland 4072, Australia}

\author{I. Pikovski}
\affiliation{University of Vienna, Faculty of Physics, Vienna Center for Quantum
Science and Technology (VCQ), Boltzmanngasse 5, A-1090 Vienna, Austria
and
Institute for Quantum Optics and Quantum Information (IQOQI),
Austrian Academy of Sciences, Boltzmanngasse 3, Vienna A-1090, Austria}

\author{M. S. Kim}
\affiliation{QOLS, Blackett Laboratory, Imperial College London, London SW7 2BW, United Kingdom}

\date{\today} 

\begin{abstract}
Utilizing the tools of quantum optics to prepare and manipulate quantum states of motion of a mechanical resonator is currently one of the most promising routes to explore non-classicality at a macroscopic scale. An important quantum optomechanical tool yet to be experimentally demonstrated is the ability to perform complete quantum state reconstruction. Here, after providing a brief introduction to quantum states in phase space, we review and contrast the current proposals for state reconstruction of mechanical motional states and discuss experimental progress. Furthermore, we show that mechanical quadrature tomography using back-action-evading interactions gives an $s$-parameterized Wigner function where the numerical parameter $s$ is directly related to the optomechanical measurement strength. We also discuss the effects of classical noise in the optical probe for both state reconstruction and state preparation by measurement.
\end{abstract}

\maketitle

\section{Introduction}

Quantum state reconstruction (QSR) of individual quantum systems is now a cornerstone of modern experimental quantum optics as it allows the complete characterization of all the complementary properties of a quantum state and can beautifully reveal non-classicality. The technique was first experimentally performed by Smithey \etal~\cite{Smithey1993} who generated the first `quantum pictures' by homodyne tomography on squeezed light to reconstruct the Wigner function and density matrix. Since this pioneering experiment, there has been an explosion of interest employing homodyne tomography to reconstruct various quantum states of light~\cite{Breitenbach1997, Lvovsky2001, Wenger2004, Ourjoumtsev2006, Neergaard2006, Parigi2007, Lvovsky2009, Bimbard2010, Yukawa2013}. Furthermore, QSR has now been performed with numerous other quantum systems such as the vibrational mode of a molecule~\cite{Dunn1995}, the motion of trapped ions~\cite{Leibfried1996, Leibfried2003}, the motion of neutral atoms in a trap~\cite{Morinaga1999}, a microwave field reflecting between the mirrors of a cavity~\cite{Deleglise2008}, a superconducting microwave resonator in an electrical circuit~\cite{Hofheinz2009}, and neutral atom spin ensembles~\cite{Fernholz2008}. Performing QSR for these systems has enabled the empirical study of decoherence~\cite{Deleglise2008}, the characterisation of quantum states for quantum information processing~\cite{Ourjoumtsev2006, Neergaard2006, Yukawa2013} and QSR continues to be a vital tool for the development of quantum memory and quantum metrology applications.

\begin{figure}[t h]
\includegraphics[width=0.7\hsize]{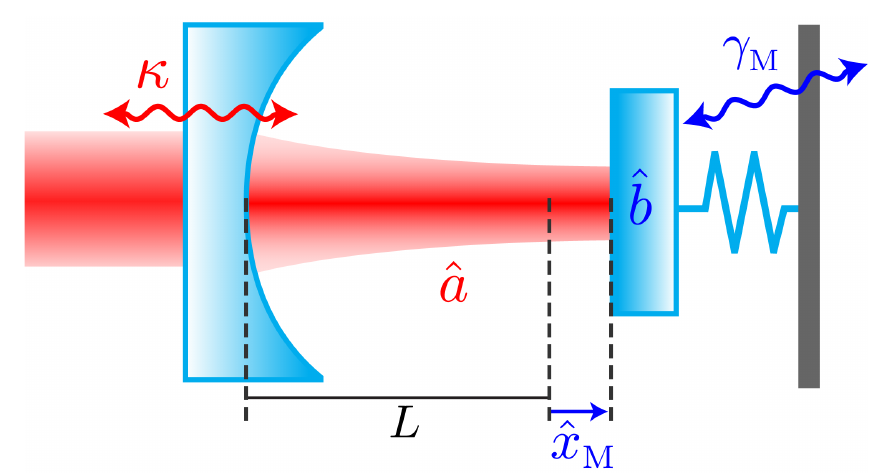}
\caption{
Schematic of a prototypical Fabry-Perot optomechanical cavity. An intracavity field $\hat{a}$ circulates inside an optical resonator with mean length $L$ and amplitude decay rate $\kappa$. One of the cavity mirrors is sufficiently small that the multiple reflections of the intracavity light---the radiation-pressure force---is sufficient to modify the mirror's momentum. The mechanically vibrating mirror may be displaced from the cavity equilibrium position by $\hat{x}_\textrm{M}$ and has decay rate $\gamma_\textrm{M}$. A primary goal of cavity quantum optomechanics is to prepare quantum states of motion of the macroscopic vibrating mirror. A key challenge that will be faced in such experiments is how to determine the quantum state of motion, which is the central topic discussed here.
}
\label{Fig:OMCavity}
\end{figure}

Cavity optomechanics is currently one of newest branches of experimental quantum optics and provides a route to explore the behaviour of macroscopic individual quantum systems. Central to this research field are light-matter interactions, typically radiation-pressure, that couple an optical field circulating inside an optical cavity to the motion of a mechanical element~\cite{Kippenberg2008, Milburn2011, Meystre2013, Aspelmeyer2013}. The prototypical cavity-optomechanical system is a Fabry-Perot cavity with one large rigid input mirror and one smaller mirror that is sufficiently compliant that even the reflection of light can modify the mirror's momentum, see Fig.~\ref{Fig:OMCavity}. With such an interaction one can then use light to both manipulate and perform precision read-out of the mechanical motion. The motivations to study optomechanical systems are numerous and include both the development of novel applications and the ability to probe the fundamental properties of nature. For example, optomechanical systems can be used for precision force sensing~\cite{Rugar1991}, bio sensing~\cite{Longo2013}, the development of hybrid quantum systems~\cite{Rabl2010}, studying open quantum system dynamics~\cite{Vanner2011}, and even probing the interface between quantum mechanics and gravity~\cite{Bose1999, Marshall2003, Pikovski2012}. The ability to perform QSR of mechanical motion will greatly assist research in each of these directions. However, an experiment demonstrating the ability to perform mechanical motional QSR is yet to be achieved.

In this article, after giving a brief introduction to the quantum description of states in phase space, we describe the various existing proposals to perform QSR of mechanical motion and discuss the parameter regimes that favour their implementation. Our presentation in this section is chosen to aid later discussion. We then present how the measurement strength of a back-action-evading measurement of a mechanical quadrature for state tomography yields a $s$-parametrised quasi-probability distribution. Also, we compute the effects of classical phase and amplitude noise in QSR and state preparation by measurement.

\section{An introduction to phase-space representations of quantum states}

Typically quantum mechanical states and operators are mathematically described as objects within a Hilbert space. However, such operators can be equivalently described using a quantum mechanical phase space \cite{Groenewold1946, Moyal1949, Hillery1997, Schleich2001, Barnett2002}. The one-to-one correspondence between operators and phase space functions can be established by a Wigner-Weyl transformation. In particular, for every density matrix one can assign a phase space quasi-probability distribution. There are several forms of quasi-probability distributions, which are however all interrelated via two-dimensional  convolution. The most common representations are the Wigner function $W(\alpha, \alpha^*)$ \cite{Wigner1932}, the Husimi $Q$-function $Q(\alpha, \alpha^*)$ \cite{Husimi1940} and the Glauber-Sudarshan $P$-function $P(\alpha, \alpha^*)$ \cite{Glauber1963, Sudarshan1963}. More generally, the $s$-parameterized quasi-probability distribution $\mathcal{P}(s,\alpha,\alpha^*)$ describing a quantum state $\hat{\rho}$ can be found from the $s$-parameterized characteristic function $C(s,\xi, \xi^*)$ \cite{Barnett2002}, which is defined as
\begin{equation}
\label{Eq:Char}
\begin{split}
C(s,\xi, \xi^*) & = \trace{\hat{\rho} e^{\xi \hat{a}^\dagger - \xi^* \hat{a}}} e^{s |\xi|^2/2}  \\
& = \trace{e^{\xi \hat{a}^\dagger} \hat{\rho} e^{-\xi^* \hat{a}}} e^{(s+1) |\xi|^2/2} \, .
\end{split}
\end{equation}
The cases $s = -1, 0 ,1$ correspond to the characteristic functions for the $Q$-function, the Wigner function, and the $P$-function, respectively. The quasi-probability distributions are then obtained via
\begin{equation}
\label{Eq:QuasiProb}
\mathcal{P}(s,\alpha,\alpha^*) = \frac{1}{\pi^2} \integ{^2\xi} C(s,\xi, \xi^*) e^{\alpha \xi^* - \alpha^* \xi } \, .
\end{equation}
In terms of the real and imaginary parts, the infinitesimal $\textrm{d}^2\xi = \textrm{d}\xi_r \textrm{d}\xi_i$, where the subscripts $r$ and $i$ denote the real and imaginary parts, respectively. The exponent then becomes $\alpha \xi^* - \alpha^* \xi= - i2(\xi_i \alpha_r - \xi_r \alpha_i)$. Thus the $s$-parameterized quasi-probability distribution can be interpreted as a 2-dimensional inverse Fourier-transform of the characteristic function, i.e. $\mathcal{F}_{(2)}^{-1}[C(s,\xi, \xi^*)]= \mathcal{P}(s,\alpha,\alpha^*)$. The quasi-probability distribution is normalized for all $s$, such that $\integ{^2\alpha} \mathcal{P}(s,\alpha,\alpha^*) = 1$.

Analysing some specific cases, for $s=-1$, one can write the trace from Eq.~(\ref{Eq:Char}) in the coherent state basis and use $\delta(x-y)=(2 \pi)^{-1} \integ{q} e^{iq(x-y)}$, which yields the $Q$-function
\begin{equation}
\begin{split}
Q(\alpha,\alpha^*)  & = \mathcal{P}(-1,\alpha,\alpha^*)  = \frac{1}{\pi} \bra{\alpha} \hat{\rho} \ket{\alpha} \, .
\end{split}
\end{equation}
The Husimi $Q$-function is a projection of the state onto a coherent state and is always positive. It can be used to directly compute anti-normal ordered correlation functions $ \mean{\hat{a}^p \hat{a}^{\dagger q}} = \trace{\hat{\rho} \hat{a}^p \hat{a}^{\dagger q} } = \pi^{-1} \integ{^2\alpha} \bra{\alpha} \hat{a}^{\dagger q} \hat{\rho} \hat{a}^p \ket{\alpha} = \integ{^2 \alpha} \,  Q(\alpha,\alpha^*) \alpha^{* q} \alpha^p$.

Substituting $s=0$ in Eq.~\eqref{Eq:QuasiProb} yields the Wigner function. To find a compact expression for the Wigner function, we first use the displacement operator property $\disd(\alpha)\hat{a} \dis(\alpha)=\hat{a}+\alpha$, such that $e^{\xi \ad - \xi^* \hat{a}} e^{\alpha \xi^* - \alpha^* \xi} = \dis(\alpha) e^{\xi \ad - \xi^* \hat{a}}  \disd(\alpha) = \dis(\alpha) \dis(\xi)  \disd(\alpha) $. The Wigner function therefore becomes $W(\alpha, \alpha^*)= \pi^{-2} \integ{^2\xi} \trace{\disd(\alpha) \hat{\rho} \dis(\alpha) \dis(\xi) } $. The resulting operator $\integ{^2\xi} \dis(\xi)$ is proportional to the parity operator $\hat{\wp}$. To see this, we can express the displacement operator in terms of the quadratures $\hat{P}=i(\ad - \hat{a})/\sqrt{2}$ and $\hat{X}=(\ad + \hat{a})/\sqrt{2}$ as $\dis(\xi) = e^{-i \xi_r \sqrt{2}\hat{P}} e^{i \xi_i \sqrt{2} \hat{X}} e^{i \xi_r \xi_i} $. Inserting identity in terms of the $\hat{P}$-quadrature eigenstates yields $\integ{^2 \xi} \dis(\xi) = \integ{^2 \xi} \integ{p} e^{i \xi_r \xi_i} e^{-i \xi_r \sqrt{2}p} \ket{p} \bra{p-\sqrt{2}\xi_i} = 2 \pi \! \integ{\xi_i} \integ{p} \,  \delta(\sqrt{2}p - \xi_i) \ket{p} \bra{p- \! \sqrt{2}\xi_i} = 2 \pi \! \integ{p} \! \ket{p} \bra{-p} = 2 \pi \hat{\wp}$. In the last step, we used the property of the parity operator $\hat{\wp} \ket{p} = \ket{-p}$. The parity operator swaps the sign of any arbitrary coherent state and can also be written as $\hat{\wp} = (-1)^{\hat{n}}$, where $\hat{n}$ is the number operator \cite{Barnett2002}. The Wigner function can then be expressed as
\begin{equation}
\label{Eq:Wfcn}
\begin{split}
W(\alpha,\alpha^*)  & = \mathcal{P}(0,\alpha,\alpha^*)  = \frac{1}{\pi^2} \integ{^2 \xi} \trace{\disd(\alpha) \hat{\rho} \dis(\alpha) \dis(\xi) } \\
& = \frac{2}{\pi} \trace{\disd(\alpha) \hat{\rho} \dis(\alpha) \hat{\wp} } \, .
\end{split}
\end{equation}
The Wigner function at a particular point $\alpha$ is therefore proportional to the expectation value of the parity operator for a state displaced by $-\alpha$. Evaluating the trace in Eq.~\eqref{Eq:Wfcn} in terms of the quadratures yields Wigner's formula~\cite{Wigner1932}
\begin{equation}
W(\alpha_r,\alpha_i) = \frac{2}{\pi} \integ{x} e^{-i 2 \sqrt{2} \alpha_i x} \bra{\sqrt{2} \alpha_r + x } \hat{\rho} \ket{ \sqrt{2} \alpha_r - x} \, .
\end{equation}
The Wigner function is always well-defined, but can have negative values. It can be used to directly compute symmetrically-ordered correlation functions, where expressions are symmetrized with respect to $\hat{a}$ and $\ad$, i.e. $\mean{(\hat{a}^{\dagger q} \hat{a}^p)_\textrm{sym}}  = \integ{^2 \alpha} \,  W(\alpha,\alpha^*) \alpha^{* q} \alpha^p $. The Wigner function also has the unique property that integrating over one quadrature yields the probability distribution for the conjugate quadrature. A straightforward computation shows that $\integ{P} \, W(X,P) = \bra{X} \hat{\rho} \ket{X} = \prob{X}$ and $\integ{X} \, W(X,P) = \prob{P}$. In this respect, the Wigner function has the closest resemblance with a classical phase-space probability distribution. Possible negative components of the Wigner function, however, have no classical analogue \cite{Hillery1997, Srinivas1975} and therefore such negativities are often used as a condition for non-classicality. All Gaussian states have a positive Wigner function.

Finally, substituting $s=1$ in Eq.~\eqref{Eq:QuasiProb} gives the Glauber-Sudarshan $P$-function, which is a common mathematical tool in quantum optics. A state $\hat{\rho}$ can be expressed in terms of the $P$-function as
\begin{equation}
\hat{\rho} = \integ{^2 \alpha} P(\alpha,\alpha^*) \ket{\alpha} \bra{\alpha} \, .
\end{equation}
This expression can be verified, using the coherent state basis for the trace in Eq.~\eqref{Eq:QuasiProb}, i.e. $\mathcal{P}(1,\alpha,\alpha^*) = \pi^{-3} \! \integ{^2\xi} \! \integ{^2\beta} \! \integ{^2\gamma} \exp[|\xi^2|+\xi(\beta^* - \alpha^*) - \xi^*(\beta - \alpha)] \abs{\braket{\beta}{\gamma}}^2 P(\gamma, \gamma^*) = \integ{^2 \gamma} \delta(\alpha_i - \gamma_i) \delta(\alpha_r - \gamma_r) P(\gamma, \gamma^*) = P(\alpha, \alpha^*)$.  The Glauber-Sudarshan $P$-function can be used to compute any normal-ordered correlation functions, i.e. $\mean{\hat{a}^{\dagger q} \hat{a}^p} = \trace{\hat{a}^p \hat{\rho}  \hat{a}^{\dagger q} } = \integ{^2\alpha} P(\alpha,\alpha^*) \alpha^{* q} \alpha^p $. However, the $P$-representation is not always positive and for pure states it is only defined through generalized functions ($\delta$-functions and derivatives thereof).  In quantum optics, non-classicality is therefore often referred to in terms of the $P$-function, i.e. whether a state can be decomposed into a mixture of coherent states. A squeezed state, for example, has no Wigner negativity, but has a highly singular $P$-function \cite{Barnett2002, Walls1983}. Some examples of quasi-probability distributions are provided later in this article, see Fig.~\ref{Fig:QuasiProb}, where we discuss the effects of imperfect quadrature measurement.

The quasi-probability distributions discussed above are interrelated via two-dimensional convolution with a Gaussian. From Eq.~\eqref{Eq:Char} it follows that the characteristic functions of the same state but for two different distributions are related via $C(s',\xi, \xi^*) = C(s,\xi, \xi^*) e^{(s'-s) |\xi|^2/2}$. The inverse-Fourier transform of this expression is the $s'$-parametrized probability distribution. Using the Fourier-convolution theorem and since $\mathcal{F}_{(2)}^{-1}[C(s,\xi, \xi^*)]= \mathcal{P}(s,\alpha,\alpha^*)$ and $\mathcal{F}_{(2)}^{-1}[e^{-(s-s') |\xi|^2/2}]=2/(\pi (s-s')) e^{-2|\alpha|^2/(s-s')}$, provided $s>s'$, the relation between the $s'$- and $s$-parametrized probability distributions is
\begin{equation}
\label{Eq:QuasiProbRelation}
\mathcal{P}(s',\alpha,\alpha^*) = \frac{2}{\pi(s-s')} \integ{^2\beta} \mathcal{P}(s,\beta,\beta^*) e^{-2|\alpha - \beta|^2/(s-s')} \, .
\end{equation}

A common technique for quantum state reconstruction relies on measurements of the marginals $M(X,\theta)= \prob{X_{\theta}}=\bra{X_{\theta}} \hat{\rho} \ket{X_{\theta}}$, which have a one-to-one relation to the quasi-probability distributions~\cite{Vogel1989}. The characteristic function for the marginals can be written as $C_m(\eta, \theta) =\trace{\hat{\rho} \, e^{i  \eta \hat{X}_{\theta}} }$, i.e., the marginals can be found from the one-dimensional inverse-Fourier transform $M(X, \theta) = \mathcal{F}^{-1}[C_m(\eta, \theta)]$. Since $\hat{X}_{\theta} = (\hat{a} e^{-i \theta} + \ad e^{i \theta})/\sqrt{2}$, the characteristic function of quasi-probability distributions, Eq.~\eqref{Eq:Char}, can be written in terms of the characteristic function of the marginals as
\begin{equation}
\label{Eq:CharMarginal}
C(s,\xi=i \eta  e^{i \theta}/\sqrt{2}) = C_m(\eta, \theta) e^{s \eta^2/4} \, .
\end{equation}
Thus, there is a direct correspondence between the marginals and the quasi-probability distributions. Writing the latter out explicitly as the inverse-Fourier transform of the characteristic function yields the relation between quasi-probability and marginal distributions \cite{Vogel1989}
\begin{equation}
\begin{split}
P(s, \xi, \xi^*) & = \frac{1}{\sqrt{2} \pi^2}  \integlim{0}{\infty}{\eta} \integlim{0}{2 \pi}{\theta} \integlim{-\infty}{\infty}{x} M(x,\theta) \times \\
 & \eta \, \exp[{s \eta^2/4+i \sqrt{2} \eta(\sqrt{2} x - \xi_i \sin\theta - \xi_r \cos\theta)}] .
\end{split}
\end{equation}
The quasi-probability distribution can then be obtained from a set of measured marginals $M(x,\theta)$ and for $s=0$, this transformation is known as the inverse-Radon transformation. There are various numerical methods for this type of inversion, see e.g. Refs.~\cite{Lvovsky2009, Lvovsky2004}, which can take into account experimental restrictions such as a finite number of marginal angles $\theta$.

\section{Current proposals for mechanical state reconstruction and experimental progress}

Prior to reviewing the various techniques that could allow QSR of mechanical motion a brief introduction to cavity optomechanics is provided. The notation and definitions introduced here and in the previous section will be used throughout the remainder of this article.

For a Fabry-Perot optomechanical cavity, as shown in Fig.~\ref{Fig:OMCavity}, the mechanical position dependent cavity resonance frequency is $\omega_\textrm{C}(x_\textrm{M}) \simeq \overline{\omega_\textrm{C}}(1-x_\textrm{M}/L)$, where $x_\textrm{M}$ is a mechanical displacement elongating the cavity, and $\overline{\omega_\textrm{C}}$ is the mean cavity resonance for mean cavity length $L$. Considering a quantized harmonic mechanical oscillator the Hamiltonian for such a system is
\begin{equation}
\label{Eq:Hamilt}
\frac{\hat{H}}{\hbar} = \omegam \bd \hat{b} + \Delta \ad \hat{a} - g_0 \ad \hat{a} (\hat{b} + \bd) + \frac{\hat{H}_\textrm{d}}{\hbar},
\end{equation}
which includes the mechanical free energy, the cavity energy, the radiation-pressure interaction, and the optical drive $\hat{H}_\textrm{d}/\hbar = \mathcal{E}^*\hat{a} + \mathcal{E}\ad$, respectively, in a frame rotating at the optical drive frequency which may be detuned from cavity resonance by $\Delta$. Here, $\hat{a}$ ($\hat{b}$) is the cavity-field (mechanical) annihilation operator, $\omegam$ is the mechanical angular frequency, $g_0 = \overline{\omega_\textrm{C}} \xzp/L$ is the optomechanical coupling rate, where $\xzp = \sqrt{\hbar/2m\omegam}$ is the zero-point motion of the mechanical mode that has effective mass $m$. 

The dynamics of the light and mechanical oscillator as they interact with one another and their respective baths are governed by the Langevin equations
\begin{equation}
\label{Eq:Motion}
\begin{split}
\dot{\hat{a}} & = -i(\Delta - g_0(\hat{b}+\hat{b}^\dagger))\hat{a} - i \mathcal{E} - \kappa \hat{a} + \sqrt{2\kappa}\hat{a}_\textrm{in} , \\
\dot{\hat{b}} & = -i\omegam \hat{b} + i g_0 \hat{a}^\dagger \hat{a} - \gamma_\textrm{M} \hat{b} + \sqrt{2\gamma_\textrm{M}}\hat{b}_\textrm{in} ,
\end{split}
\end{equation}
where $\gamma_\textrm{M}$ is the mechanical amplitude decay rate and the input noise terms have been introduced, which include both the ubiquitous quantum noise and any classical noise present. Here, a particular Brownian noise model of the mechanical motion has been used where the mechanical damping affects both of the mechanical quadratures equally. We would like to note that all the richness of cavity quantum optomechanics stems from subtle variations of these equations of motion and the type of measurement performed on the light field.

Common to many of the QSR schemes discussed below is the use of light to probe the mechanical degrees of freedom. This is achieved via cavity drive and then measurement of the light that decays out of the optomechanical cavity. The light field leaving the cavity is determined via the input-output relation $\hat{a}_\textrm{out} = \sqrt{2\kappa}\hat{a} - \hat{a}_\textrm{in}$, see Ref.~\cite{Gardiner1985}, which has recently been characterised for Gaussian processes~\cite{Tufarelli2012}.

\subsection{State transfer}

One technique to perform QSR of mechanical motion is to first transfer the mechanical quantum state onto a light field and then perform optical homodyne tomography. This technique has the advantage that it can utilise existing optical homodyne tomography techniques. Moreover, being able to achieve quantum state transfer between the mechanical motion and light, and vice-versa, is an important goal in its own right as this amounts to the implementation of a continuous-variable quantum memory.

Building upon prior theoretical work in cavity quantum electrodynamics, which considered state transfer between light and the motion of a trapped atom~\cite{Parkins1999}, a technique for optomechanical quantum state transfer was proposed based on detuned cavity drive~\cite{Zhang2003}. To understand this technique quantitatively we return to Eq.~(\ref{Eq:Motion}). First, a displaced frame is entered that follows the mean of the optical and mechanical field operators ($\hat{a} \rightarrow \hat{a} + \alpha$, and $\hat{b} \rightarrow \hat{b} + \beta$) and the equations of motion are linearised. Next, a rotating frame is entered, i.e. $\hat{a} \rightarrow \hat{a} e^{-i\Delta t}$, and $\hat{b} \rightarrow \hat{b}e^{-i\omegam t}$ and the equations of motion become
\begin{equation}
\label{Eq:MotionFrame}
\begin{split}
\dot{\hat{a}} & \simeq i g_0 \alpha (\hat{b}e^{-i(\omegam - \Delta)t} + \hat{b}^\dagger e^{i(\omegam + \Delta)t}) -\kappa \hat{a} + \sqrt{2\kappa}\hat{a}_\textrm{in} , \\
\dot{\hat{b}} & \simeq i g_0 (\alpha^* \hat{a} e^{-i(\Delta - \omegam)t} + \alpha \hat{a}^\dagger e^{i(\Delta + \omegam)t}),
\end{split}
\end{equation}
where the mechanical damping and noise terms have been dropped for clarity.

To achieve a state-transfer interaction one sets $\Delta = +\omegam$, that is, drive on the red sideband. Provided that the cavity optomechanical system operates in the resolved sideband regime, i.e. $\kappa \ll \omegam$, the terms that oscillate at $2\omegam$ can be neglected by making a rotating-wave approximation. In this regime, the field operator $\hat{a}$ gets correlated to $\hat{b}$ and vice-versa, which is described by the so-called beam-splitter Hamiltonian, $\hat{H}_\textrm{int} \propto \hat{a} \hat{b}^\dagger + \hat{a}^\dagger \hat{b}$. (As a side remark, it should also be noted that by setting $\Delta = -\omegam$ in Eq.~(\ref{Eq:MotionFrame}), i.e. drive on the blue sideband, the unitary dynamics are governed by a two-mode-squeezing Hamiltonian, $\hat{H}_\textrm{int} \propto \hat{a} \hat{b} + \hat{a}^\dagger \hat{b}^\dagger$. This interaction generates optomechanical entanglement~\cite{Mancini2003, PalomakiEnt2013} and can be used for conditional phonon-level operations~\cite{Lee2012, VannerPRL2013}.) By driving this beam-splitter interaction, in the absence of decoherence, the mechanical and cavity field states swap at the rate of $2g_0\alpha$ \cite{Parkins1999, Zhang2003}. In order to achieve an efficient quantum state transfer it is necessary that this rate exceed both the optical cavity decay rate~\cite{GrobNat2009} and the mechanical decoherence rate, which has recently been achieved for both optomechanical~\cite{Verhagen2012} and electromechanical systems~\cite{OConnell2010, Teufel2011}.

The beam-splitter interaction is commonly employed in optomechanics experiments as this is the interaction behind sideband cooling~\cite{Braginsky1970, Schliesser2009, Groblacher2009, TeufelCool2011, Chan2011}, which is also utilized for laser cooling of trapped ions~\cite{Diedrich1989}. Here, sideband cooling can be viewed as a partial state transfer between the low-entropy optical state and the mechanical state. This beam-splitter interaction has also been considered to characterise mechanical squeezing generated by external drive~\cite{Woolley2008, Seok2012}.

Both optomechanical~\cite{Verhagen2012} and electromechanical~\cite{Palomaki2013} coherent quantum state transfer are becoming a practical reality. The key challenge that will be faced in future experiments in this direction is how to achieve a high efficiency. This is particularly important for mechanical QSR via state transfer to light as, even in the limit of perfect optical detection efficiency, state transfer inefficiency is equivalent to convolving the mechanical Wigner function with a Guassian. Such a convolution, see Eq.~\eqref{Eq:QuasiProbRelation}, takes a Wigner function `towards' a $Q$-function and thus any possible negativities can be washed away. This effect is well known in quantum optics where loss prior to optical homodyne tomography yields an $s$-parametrised Wigner function where the $s$-parameter is related to the loss parameter~\cite{Leonhardt1993, Lutkenhaus1995}.

\subsection{Characteristic function}

Another technique to fully quantify a quantum state is to measure the characteristic function of the Wigner quasi-probability distribution, Eq.~(\ref{Eq:Char}). Early theoretical work in this direction was reported in Refs.~\cite{Wilkens1991, Kim1998} where it was shown that the interaction between an optical cavity field and an atom, can be used to directly determine the characteristic function of the optical field. In the scheme of Ref.~\cite{Wilkens1991}, two modes $\hat{a}_1$ and $\hat{a}_2$ of the optical field are used,  which interact with an atom via the Jaynes-Cummings Hamiltonian $\hat{H}/\hbar = g (\hat{\sigma}_+ (\hat{a}_1 + \hat{a}_2) + \hat{\sigma}_- (\hat{a}_1^{\dagger} + \hat{a}_2^{\dagger}))$. After an interaction time $t$ the atom leaves the cavity and the population of its excited state is measured. The second optical mode can act as a local oscillator with phase difference $\phi$ where $\langle \hat{a}_2^{\dagger} \hat{a}_2 \rangle \gg \langle \hat{a}_1^{\dagger} \hat{a}_1 \rangle$. In this regime, the probability to find the atom excited is $\frac{1}{2} (1 + K C(\mu) + K^* C^*(\mu) )$, where $C(\mu)$ is the characteristic function for the Wigner function of the first optical mode with $\mu = igt e^{i\phi}$ and where $K$ depends on the interaction strength and the initial population difference of the two atomic states. Thus the measurement of the atom for varying interaction times and local oscillator phase can reveal the characteristic function of the optical field. In Ref.~\cite{Kim1998} it was shown that a similar result can be obtained without the use of a second optical mode, but with a displacement operation acting on the optical field. When probed by an atom, the displaced optical field induces a population inversion of the atom that is again related to the Wigner characteristic function of the initial optical mode. 

Building upon these prior works, Ref.~\cite{Singh2010} showed theoretically that the characteristic function of a mechanical resonator state can be determined by coupling it magnetically to the hyperfine levels of an atom. On resonance and in the rotating wave approximation, it was shown that the coupling reduces to the Jaynes-Cummings model. Using an additional optical field to induce Raman transitions in the atom, the mechanical Wigner characteristic function can be obtained in a similar manner to Ref.~\cite{Wilkens1991}. The Raman field effectively acts as a local oscillator and the Wigner characteristic function of the mechanical modes can be determined from the population inversion of the atom.

\subsection{Displacement and number measurement}

An important tool throughout quantum optics is the ability to measure the number of energy quanta. Knowing the number distribution for a harmonic oscillator provides significant information about the state but it does not quantify the strength and phases of any quantum coherences between different excitations. Together the number distribution and coherences fully describe the quantum state and can be represented with the density matrix in the number basis. In practice one is generally able to readily measure the number distribution, the diagonal elements of the density matrix, and to access the coherences one can first displace the quantum state. Via Eq.~(\ref{Eq:Wfcn}) we also see that this procedure, of first displacing the state and then performing a number measurement, provides a route to determine the Wigner function~\cite{Wallentowitz1996, Banaszek1996}, where the number measurement outcomes are used to determine the expectation of the parity operator. Using this type of procedure the quantum motional state of a trapped ion was reconstructed~\cite{Leibfried1996}. Also, in a cavity-QED context, using a Ramsey interferometer to measure microwave cavity photon number following a displacement operation, non-classical cavity field states were reconstructed~\cite{Deleglise2008}. And, the Wigner function of non-classical microwave states in a coplanar waveguide resonator was determined by photon number measurements made by coupling to a Josephson phase qubit after a displacement operation to the resonator~\cite{Hofheinz2009}. It should also be noted here that this technique of determining the Wigner function does not require numerical inversion from a set of complementary measurements. However, the technique works over a limited range of displacements and can thus be used to reconstruct states with features close to the origin of phase space.

In principle, this procedure can be applied to reconstruct the motion of a mechanical resonator as both displacement operations and phonon number measurements are available in quantum optomechanics. Indeed, in a recent electromechanics experiment a micromechanical bulk dilatational resonator was coupled to a Josephson phase qubit~\cite{OConnell2010}. Provided sufficient coherence times of the mechanical oscillator and the phase qubit, this arrangement allows measurement of the mechanical Wigner function in a similar manner to Ref.~\cite{Hofheinz2009}.

\subsection{Mechanical quadrature measurement}

Studying the position or momentum distributions of a quantum system is an important aspect of numerous experiments. As a prominent example, these distributions are routinely studied in experiments with ultracold atomic gases~\cite{Bloch2008, Stamper2012}, via e.g. absorption imaging or Bragg spectrscopy~\cite{Stenger1999, Rolston2002}. Furthermore, sideband spectroscopy is a vital tool to study the motion of trapped ions~\cite{Diedrich1989}. Building upon these previous techniques it has recently been theoretically proposed how to perform Wigner reconstruction of the motion of trapped particles~\cite{Isart2011} and large molecules in diffraction experiments~\cite{LeeSK2012}.

In cavity optomechanics the radiation-pressure interaction, Eq.~(\ref{Eq:Hamilt}), naturally provides a coupling to the mechanical position and a number of mechanical QSR schemes aim to exploit this interaction for mechanical QSR via quadrature tomography~\cite{Vogel1989}. There has been considerable progress made in optomechanics with high-precision continuous monitoring of the mechanical position and classical phase-space measurements~\cite{Rugar1991, Blair1995, Tittonen1999, Briant2003, Abbott2009}. However, of vital importance for mechanical QSR is to be able to resolve features smaller than the width of mechanical zero-point motion, which is not possible with a continuous strength position measurement due to the standard quantum limit~\cite{BraginskyBook}. In order to circumvent this limit it is necessary to use quadrature amplification~\cite{Rugar1991, Ahmad2000} or a back-action-evading (BAE) measurement~\cite{Braginsky1975, Thorne1978, Unruh1979, Braginsky1996, Grangier1998}. For the case of optomechanics, from Eq.~(\ref{Eq:Hamilt}), it can be seen that the interaction Hamiltonian does not commute with the full Hamiltonian and so the back-action noise imparted to the mechanical momentum evolves into position noise. While increasing the optical probe power decreases the optical phase uncertainty it increases the back-action noise on the mechanics due to the optical number-phase uncertainty product $\Delta n \Delta \theta \geq 1/2$. These effects combined are the origin of the standard quantum limit that prevents mechanical displacements smaller than the ground-state width to be resolved. BAE measurement techniques were first experimentally implemented to measure the quadrature amplitudes of optical fields~\cite{Levenson1986, Porta1989} and are now also an important technique for atomic ensemble spin squeezing experiments~\cite{Kuzmich2000, Teper2008, Takano2009, Appel2009, SchleierSmith2010, Sewell2013}. Common to all optomechanical BAE measurement techniques is a time-varying measurement strength. In the remainder of this section we will describe the three main optomechanical BAE measurement techniques and how they can be used for mechanical QSR.

\subsubsection{Two-toned drive}

The basic principle of the `two-toned-drive' BAE measurement technique is to use a time varying intracavity amplitude $\alpha(t) = \alpha_0 \cos(\omegam t)$ to probe not the position of the oscillator as a function of time but rather the slowly varying quadrature $\hat{X}_{1}$ defined via $\hat{X}_\textrm{M}(t) = \hat{X}_1 \cos(\omegam t) + \hat{X}_2 \sin(\omegam t)$. Linearising the radiation pressure interaction about this oscillating cavity amplitude, and going into the mechanical rotating frame, the intracavity Hamiltonian becomes $\hat{H}/\hbar \rightarrow -g_0 \alpha_0 (\hat{a} + \hat{a}^\dagger) [\hat{X}_1 (1 + \cos(2\omegam t)) + \hat{X}_2 \sin(2\omegam t)]/\sqrt{2}$ and it can be seen that, over a time scale long compared to $\omegam^{-1}$, which is achieved by operating in the resolved sideband regime where $\kappa \ll \omegam$, the light accumulates a phase proportional to $\hat{X}_1$. It must be noted that even though $[\hat{X}_1, \hat{X}_2] = i$ these conjugate quadratures are not coupled via mechanical harmonic motion and the measurement technique causes the radiation-pressure back-action noise to be added only to the mechanical quadrature $\hat{X}_2$. This technique can then be used to surpass the SQL as increasing the optical probe strength does not perturb $\hat{X}_1$.
Such a time-varying intracavity amplitude can be generated using an amplitude modulated external resonant drive. Akin to Eq.~(\ref{Eq:Motion}), the equation of motion for the intracavity amplitude is $\dot{\alpha} = -i \mathcal{E} - \kappa \alpha$. By setting $\mathcal{E} = -i \mathcal{E}_0 \sin(\omegam t)$ one obtains $\alpha(t) \simeq (\mathcal{E}_0 \omegam/(\kappa^2 + \omegam^2))\cos(\omegam t)$ for $\kappa \ll \omegam$.

The two-toned-drive technique obtains its name from the form of the external drive, which in Fourier space has tones at $\pm \omegam$, and was first proposed by Braginsky and colleagues~\cite{Braginsky1980, BraginskyBook} to surpass the SQL and improve force sensing. Recently, a quantum noise analysis of this technique was performed for conditional squeezing of mechanical motion by measurement~\cite{Clerk2008}. The experimental state-of-the-art using this technique is performed with electromechanical systems, where a nano-mechanical resonator is capacitively coupled to a superconducting microwave resonator driven with a modulated input. Using this arrangement a sensitivity very close to the mechanical zero-point extension was achieved~\cite{Hertzberg2010} and more recently an experiment was performed that both observed the quantum back-action noise and then demonstrated its evasion using a two-toned drive~\cite{Suh2013}.

Being a BAE technique, the two-toned-drive allows a sub-SQL measurement of the mechanical position probability distribution that can reveal features smaller than the ground-state width. Mechanical quantum state tomography can then be performed using the two-toned-technique by changing the phase of the probe field, i.e. $\alpha(t) = \alpha_0 \cos(\omegam t - \phi)$, so that a rotated slowly varying mechanical quadrature $\hat{X}_\phi = \hat{X}_1 \cos(\phi) + \hat{X}_2 \sin(\phi)$ is measured. Obtaining the position probability distribution for several angles $\phi$ can then be used to determine the mechanical Wigner function via the inverse Radon transformation. Further detail of this kind of tomography protocol will be provided in sections to follow. We would also like to note that a two-toned laser drive has been proposed to couple the motion of a trapped ion to its electronic state for QSR via measurement of the vibrational quadrature marginals~\cite{Wallentowitz1995}.

\subsubsection{Variational-output interferometry} 

Another technique to circumvent the SQL is to use a continuous optomechanical interaction but a time or frequency dependent measurement on the output signal beam. This type of `variational-output interferometer' was first considered by Vyatchanin, Matsko, and Zubova~\cite{Vyatchanin1994, Vyatchanin1995, Kimble2001}. They proposed that a frequency dependent homodyne phase, i.e. measurement of a frequency-varying quadrature of the optical field, can be used to perform sub-SQL measurements of the mechanical position to better infer variations in a classical force acting on the oscillator. They also considered using a modulated local oscillator phase so that the optical phase and amplitude quadrature are measured alternately in time. In this arrangement, the homodyne photocurrent gives information about the back-action noise imparted to the mechanical oscillator when the optical amplitude quadrature is measured and the oscillator's position when the optical phase quadrature is measured. By optimally processing this time-dependent photocurrent~\cite{Miao2010}, it is possible to estimate the mechanical position to a precision below the SQL and hence be used for quantum state tomography of the mechanical motion.

\subsubsection{Pulsed and stroboscopic} 
\label{Sec:Pulse}

In the late 1970s, at about the same time as the two-toned-drive technique was conceived, Braginsky and colleagues proposed a pulsed optomechanical interaction for BAE measurements of the mechanical position for weak force sensing applications~\cite{Braginsky1978}. By using a pulse of light with duration much shorter than a mechanical period of motion the back-action imparted to the mechanical momentum quadrature has insufficient time to evolve into position noise. By then increasing the optical probe strength the position can be estimated with higher precision limited by the energy in the coherent optical pulse and not the SQL. Such a pulsed interaction can be repeated periodically in synchrony with the mechanical motion, i.e. stroboscopically, which is analyzed in Ref.~\cite{Braginsky1978Strob, Danilishin2002}. For the cavity to accommodate a broadband pulse and allow a build-up and decay of the cavity field much faster than the mechanical motion, $\kappa \gg \omegam$ is required. This is the opposite regime to the two-toned-drive technique, which requires the sideband resolved regime. As different physical realisations of cavity optomechanical systems naturally operate in one regime or the other BAE position measurement is then available by choosing between the pulsed or two-toned technique. 

More recently such a pulsed interaction was utilised in a proposal for low-entropy and quantum-squeezed state preparation of mechanical motion by measurement~\cite{Vanner2011}. In this regime of pulsed quantum optomechanics, a key capability is that the mechanical squeezed position variance is limited by the pulsed position measurement precision and not by the initial mechanical thermal occupation. In contrast to the two-toned-drive technique the pulsed approach allows a measurement to be performed over a much shorter time-scale. This has the advantage that, provided sufficiently large measurement strength, a quantum squeezed state can be prepared even for a large mechanical decoherence rate. That is, the state preparation and read-out can be performed on a time-scale shorter than the mechanical decoherence time-scale, which may be shorter than the mechanical period. Another important feature allowed by using pulsed quantum measurement is that mechanical dynamics and thermal bath coupling can be probed~\cite{Vanner2011}, which allows the parameters of phenomenological master equation describing the open quantum system dynamics to be determined~\cite{Bellomo2009}.

By appropriately timing when the pulsed measurement is performed any mechanical quadrature can be measured and hence mechanical QSR can be performed. More specifically, the protocol~\cite{Vanner2011} for pulsed mechanical QSR is: (i) prepare the mechanical motional state of interest at a known time, (ii) allow the mechanical oscillator to undergo free harmonic evolution for a time $\tau = \theta/\omegam$, (iii) perform a BAE pulsed measurement and record the measurement outcome, (iv) repeat steps (i)-(iii) several times for each $\theta$ and for many $\theta$ in order to well sample several mechanical marginals. This type of mechanical tomography is central to the present work and will be discussed in the following section.

Early experimental progress using this pulse technique has recently been made and tomography of the motional state of a cantilever was performed for state reconstruction~\cite{Vanner2013}. The experiment followed the steps outlined above to obtain histograms of the mechanical marginals and used the inverse-Radon transformation to reconstruct the phase-space distribution. The position measurement had a sensitivity far from being able to resolve the mechanical ground-state width, however, the optical quantum-noise-limited measurement was able to resolve features well below the mechanical thermal state width. Indeed, the cantilever oscillator had an initial rms thermal state width of 1.2~nm, which was reduced by measurement to a conditional width of 19~pm. This corresponds to 36~dB of thermal noise squeezing in the position variance. Such a pulsed BAE interaction~\cite{Braginsky1978, Vanner2011, Vanner2013} has also been utilised in theoretical proposals to probe tunneling of a mechanical oscillator in a double-well potential~\cite{Buchmann2012}, perform strong position-squared measurements~\cite{VannerPRX2011}, generate optomechanical geometric phases~\cite{Pikovski2012, Khosla2013}, and can be used for the development of QND interfaces~\cite{Marek2010}.

\section{Reconstructing mechanical non-classicality}

With experiments now demonstrating early signs of mechanical quantisation~\cite{OConnell2010, Lee2011, Lee2012} it is an important time in cavity optomechanics to turn to experimental phase-space reconstruction. A phase-space quasi-probability distribution completely characterises the motional quantum state and any negativity is a sufficient and unambiguous signature of non-classicality. The methods for mechanical QSR we have reviewed here are based on the approximate linearised optomechanical interaction, but as experiments push into the non-linear regime of optomechanics, where single photon radiation pressure is significant, it has been proposed that mechanical QSR can be performed by analysing the photon emission spectra~\cite{Liao2014}. Having now completed our review of how to perform mechanical QSR and obtain the quantum phase-space distribution we now discuss how the indirect measurement of the mechanical position with an optical probe, which has intrinsic optical quantum noise and possible classical phase noise, affects the reconstructed mechanical phase-space distribution. Then, we discuss how the presence of classical noise on the optical probe affects state conditioning, which is intimately related to quadrature measurement and tomography. These results are readily applicable to other marginal measurement techniques.

\subsection{Realistic mechanical state tomography}

We model the presence of classical phase and amplitude noise on the coherent probe with use of the $P$-function, i.e. $\hat{\rho}_\textrm{L} = \int \textrm{d}^2\alpha P(\alpha, \alpha^*) \ketbra{\alpha}{\alpha}$. We consider the experimentally relevant case of a Gaussian $P$-function that can have asymmetric phase and amplitude noise. Expressed in terms of the optical quadratures $X_\alpha = \sqrt{2}\textrm{Re}(\alpha)$, and $P_\alpha = \sqrt{2}\textrm{Im}(\alpha)$, we use
\begin{equation}
P(X_\alpha, P_\alpha) = \frac{1}{2\pi\sigma_\textrm{X} \sigma_\textrm{P}} \exp \left[ \frac{-(X_\alpha - \overline{X}_\textrm{L})^2}{2\sigma_\textrm{X}^2} - \frac{P_\alpha^2}{2\sigma_\textrm{P}^2}\right] ,
\end{equation}
where $\overline{X}_\textrm{L}$ corresponds to the mean real amplitude, and $\sigma_\textrm{X,P}$ describe the widths of the amplitude and phase noise, respectively.

\begin{figure}[h]
\includegraphics[width=1.0\hsize]{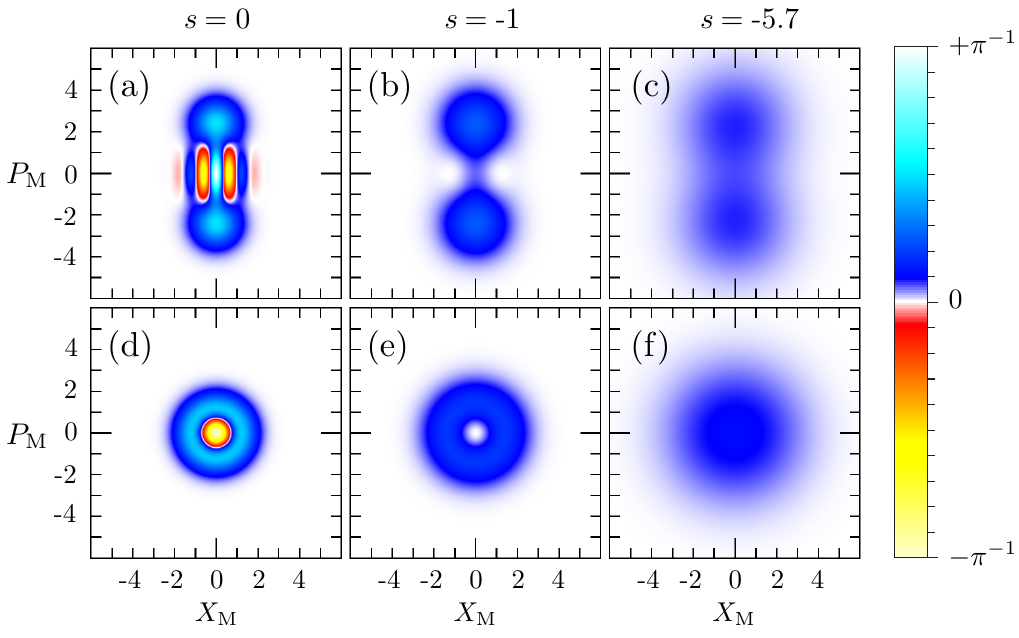}
\caption{Quasi-probability distributions for different levels of mechanical quadrature measurement strength. In the first row the state $\ket{\beta} + \ket{-\beta}$ for $\beta = 1.7i$ is shown for: (a) $s = -(1+2\sigma^2_\textrm{P})/\chi^2 = 0$, which corresponds to the Wigner function; (b) $s = -1$, which corresponds to the $Q$-function; (c) $s=-5.7$. In the second row a mechanical Fock state $\ket{1}$ is shown for: (d) $s=0$; (e) $s=-1$; (f) $s=-5.7$. Note that there are two regimes of `resolvability', i.e. in plots (a) and (d) the negativity is present, which is not seen in the other plots, and in (b) and (e) the structure of the states is seen, i.e. the two peaks and the ring, which is not well resolved in (c) and (f) as the additional smoothing causes the reconstructed states to appear more Gaussian.
}
\label{Fig:QuasiProb}
\end{figure}

The mechanical state to be reconstructed $\hat{\rho}_\textrm{M}^\textrm{in}$, is allowed to evolve freely for phase-space angle $\theta$ before interacting with a probe pulse of light. The pulse of light is sufficiently short that the interaction may be modelled by the unitary $\hat{U} = e^{i\lambda \hat{a}^\dagger \hat{a} \hat{X}_\textrm{M}}$, where the mechanical harmonic motion and mechanical open system dynamics during the short pulsed interaction are neglected. The correlated light-matter bipartite state immediately after the interaction is $\hat{\rho}_\textrm{LM} = \hat{U} \hat{\rho}_\textrm{M}^\textrm{in}(\theta) \hat{\rho}_\textrm{L} \hat{U}^\dagger$. To infer the mechanical quadrature a time-domain homodyne measurement of the phase quadrature of the outgoing pulse is performed. We model this measurement as a projection onto $\bra{\PL}$ and we note that any electronic noise associated with this measurement may be subsumed into the optical phase noise $\sigma_\textrm{P}$. It is now useful to introduce the measurement operator $\hat{\Upsilon}_\alpha = \bra{\PL}\hat{U}\ket{\alpha}$, which allows one to conveniently compute the homodyne probability distribution and the mechanical conditional state following a measurement. Note that $\hat{\Upsilon}_\alpha$, also called a Kraus operator, is an operator that acts on the mechanical sub-space as the inner product is taken over the light. For a particular input coherent state taken from $P(\alpha, \alpha^*)$ the measurement operator is 
\begin{equation}
\begin{split}
\hat{\Upsilon}_\alpha \simeq \pi^{-1/4}\exp[-{\textstyle\frac{1}{2}}(\PL - P_\alpha - \chi \hat{X}_\textrm{M})^2] \times \\\exp[i \Omega \hat{X}_\textrm{M} + i \chi \delta X_\alpha \hat{X}_\textrm{M}] \,,
\end{split}
\end{equation}
where $\chi = \sqrt{20 N}g_0/\kappa$, $N$ is the mean photon number per pulse, $\delta X_\alpha = X_\alpha - \overline{X}_\textrm{L}$, and $\Omega = {\textstyle\frac{3}{\sqrt{2}}}{\textstyle\frac{g_0}{\kappa}}N$, see Ref.~\cite{Vanner2011} for details on the numerical pre-factors. Here, $\hat{\Upsilon}_\alpha$ has been linearised so that it is a Gaussian operation, which requires $\lambda^2 \sigma_{\XM}^2 \ll 1$ and $2\lambda^2\sigma_\textrm{X}^2 \sigma_{\XM}^2 \ll 1 + 2 \sigma_\textrm{P}^2$. Following the optomechanical interaction the probability distribution for obtaining homodyne outcome $\PL$ is
\begin{equation}
\label{Eq:ProbPL}
\begin{split}
\prob{\PL} = \integ{\XM} \! \integ{^2\alpha} P(\alpha,\alpha^*) \Upsilon^\dagger_\alpha(\XM) \Upsilon_\alpha(\XM) \times \\
\bra{\XM} \hat{\rho}_\textrm{M}^\textrm{in}(\theta)\ket{\XM}.
\end{split}
\end{equation}
Note that this expression depends on the positive-operator valued measure (POVM) element $\hat{\Upsilon}^\dagger_\alpha \hat{\Upsilon}_\alpha$. For our specific case, the probability distribution $\prob{\PL}$ is the mechanical marginal of interest convolved with a Gaussian proportional to $\exp[-\chi^2(\PL/\chi - \XM)^2/(1+2\sigma^2_\textrm{P})]$. This Gaussian kernel in the convolution can be interpreted as the conditional probability distribution for $\PL$ given a mechanical position eigenstate $\ket{\XM}$. As the measurement strength $\chi$ increases and the classical phase noise $\sigma_\textrm{P}$ decreases this distribution is a more faithful measurement of the mechanical marginal in the scaled outcome $\PL/\chi$. By taking the Fourier transform of Eq.~(\ref{Eq:ProbPL}), using the $P$-function and measurement operator from above, one obtains a product of the mechanical marginal characteristic function $C_m(\eta, \theta)$ with a Gaussian. Using now Eq.~(\ref{Eq:CharMarginal}) from our introduction we see that realistic mechanical state tomography allows a reconstruction of the $s$-parametrized Wigner function with
\begin{equation}
\label{Eq:sParam}
s = \frac{-(1+2\sigma_\textrm{P}^2)}{\chi^2} \,.
\end{equation}
In the limit of large $\chi$ and low phase noise, $s$ approaches zero and the quasi-probability distribution reconstructed will be close to the Wigner function. Of course, as the $s$-parametrized Wigner function uniquely represents a state and $s$ can be accurately experimentally determined then the mechanical motional state is fully characterised. It is also quite interesting to note that, while it is well known that optical loss prior to optical homodyne tomography gives a $s$-parametrized Wigner function~\cite{Leonhardt1993}, this quite different deleterious effect of imperfect measurement also results in such a distribution. We would also like to note that with knowledge of $s$ it is in principle possible to perform deconvolution to obtain a Wigner function, however, direct deconvolution is a numerically unstable procedure and is not practical in most cases~\cite{DeconvFootnote}. In light of these points, in order to see any possible negativity in the reconstructed quasi-probability distribution, one requires $s > -1$ (as the $Q$-function is positive), which amounts to
\begin{equation}
\chi^2 > 1 + 2\sigma^2_\textrm{P} \, .
\end{equation}
In the quantum noise units we are working with here, the classical noise $\sigma^2_\textrm{P}$ scales linearly with photon number. Then, as $\chi^2$ is also linearly proportional to the photon number, one can only significantly decrease $s$ by increasing the photon number provided that $2\sigma^2_\textrm{P} \lesssim 1$. To see the effects of decreasing $s$-parameter, see Fig.~\ref{Fig:QuasiProb}.

\subsection{Realistic conditioning}

Performing a BAE measurement not only allows to perform quantum state tomography of the mechanical motion but it can also be used for quantum squeezed state preparation via measurement. Thus a single tool forms a complete experimental framework for quantum state preparation and quantum state reconstruction~\cite{Vanner2011}. Using the pulsed interaction and classical noise model of above, the mechanical state of motion conditioned on the homodyne outcome is
\begin{equation}
\hat{\rho}_\textrm{M}^\textrm{out} =  \frac{\integ{^2\alpha}P(\alpha, \alpha^*)\hat{\Upsilon}_\alpha \hat{\rho}_\textrm{M}^\textrm{in} \hat{\Upsilon}_\alpha^\dagger}{\prob{\PL}} \,.
\end{equation}
For the Gaussian measurement operator $\hat{\Upsilon}_\alpha$ and $P$-function considered here, the mechanical conditional state remains Gaussian and has position mean and variance
\begin{equation}
\mean{\hat{X}_\textrm{M}} = \frac{\chi \PL}{\chi^2 + \frac{1 + 2\sigma^2_\textrm{P}}{1+2\nbar}}, \,\,\, \textrm{and} \,\,\,\, \sigma^2_{\XM} = \frac{1}{2} \frac{1 + 2\sigma^2_\textrm{P}}{\chi^2 + \frac{1 + 2\sigma^2_\textrm{P}}{1+2\nbar}} ,
\end{equation}
respectively, where $\nbar$ is the mean thermal occupation when in thermal equilibrium. Provided that $(1+2\sigma^2_\textrm{P})/(1+2\nbar) \ll \chi^2$, i.e. the large thermal occupation limit, $\mean{\hat{X}_\textrm{M}} \simeq \PL/\chi$ and $\sigma^2_{\XM} \simeq (1+2\sigma^2_\textrm{P})/2\chi^2$. This latter expression is inversely proportional to a `quantum signal-to-noise ratio', which relates the amount of signal on the light from the mechanical quantum noise to the optical phase noise. Also note that this conditional width is directly proportional to the $s$-parameter, Eq.~(\ref{Eq:sParam}), which illustrates the connection between the measurement-based state preparation and the quadrature tomography.

One can implement `cooling-by-measurement' by making a pulsed measurement, waiting for one-quarter of a period of mechanical motion, and then performing another pulsed measurement. The resulting state is Gaussian and can have a significantly decreased entropy as information about both the position and momentum has been gained. The effective thermal occupation after such measurement protocol, in the large initial thermal occupation limit, including noise from the thermal bath during the mechanical evolution is determined via
\begin{equation}
1 + 2\nbar_\textrm{eff} \simeq \sqrt{\left(\frac{1 + 2\sigma^2_\textrm{P}}{\chi^2}\right)\!\left(\frac{1 + 2\sigma^2_\textrm{P}}{\chi^2} + \frac{\nbar \pi}{Q} + \chi^2(1 + 2\sigma^2_\textrm{X})\right)} ,
\end{equation}
where $Q$ is the mechanical quality factor. Provided that $\chi^2 \gtrsim 1 + 2\sigma^2_\textrm{P}$ and the amplitude noise is small, i.e. $\sigma^2_\textrm{X} \lesssim 1/2$ this protocol can prepare a high purity mechanical state of motion with considerable resilience against high initial occupation and thermal bath coupling.

\section{Conclusions}

The ability to perform quantum state reconstruction of mechanical motion will be an invaluable tool in quantum optomechanics as the research field begins to explore quantum mechanical behaviour. There are a number of different mechanical state reconstruction techniques proposed, which are reviewed here, that suit different parameter regimes and hence different physical implementations. A key challenge of all schemes will be to perform state reconstruction with high precision. We have discussed that performing an indirect measurement of the mechanical marginals, with a back-action-evading interaction with an auxiliary probe, results in a $s$-parametrised quasi-probability distribution, where the $s$-parameter is related to the measurement strength. This results in a smoothing of the Wigner quasi-probability distribution, which reduces or can even eliminate any negativity in the quasi-probability distribution. We have also analysed how classical phase and amplitude noise affects tomography and Gaussian state preparation via measurement.

\section{Acknowledgements}
M.R.V. and I.P. acknowledge the kind hospitality provided by M.S.K. at Imperial College London where some of this research was performed. We would like to acknowledge useful discussion with Alex Szorkovszky. This work was supported by an ARC Discovery Project (DP140101638). M.S.K. would like to thank support provided by the EPSRC. I.P. would like to thank support provided by the Austrian Science Fund FWF (DK CoQuS W 1210, Individual Project 24621) and the Vienna Center for Quantum Science and Technology (VCQ).



\begin{thebibliography}{99}
\bibitem{DeconvFootnote} The numerical instability of direct deconvolution can be understood with use of the convolution theorem $\mathcal{F}[f*g(t)] = \sqrt{2\pi} \tilde{f}(\omega) \tilde{g}(\omega)$. The signal of interest $f(t)$ can in principle be determined by dividing by $\tilde{g}(\omega)$ and then performing an inverse Fourier transform. The instabilities arise during the division stage as estimates of $\tilde{g}(\omega)$ can approach zero for certain spectral components and hence lead to divergences. If the spectral properties of the signal and any additional noise are known in advance one may instead perform Wiener deconvolution, which is more robust against such instabilities. With a finite signal-to-noise ratio the deconvolution will never be perfect and hence one will still obtain an $s$-parameterised Wigner function as discussed here.
\end{thebibliography}
\end{document}